\documentclass[traditabstract]{aa} 
\usepackage{graphicx}
\usepackage{txfonts}
\begin{document}
   \title{SOAP\thanks{The tool is available at http://www.astro.up.pt/soap}}

   \subtitle{A tool for the fast computation of \\ photometry and radial velocity induced by stellar spots}

   \author{I. Boisse
          \inst{1}
          \and
          X. Bonfils\inst{2}
          \and
          N. C. Santos\inst{1,3}
          }

   \institute{Centro de Astrof\'isica, Universidade do Porto, Rua das Estrelas, 4150-762 Porto, Portugal\\
              \email{Isabelle.Boisse@astro.up.pt}
         \and
             UJF-Grenoble 1 / CNRS-INSU, Institut de PlanŽtologie et d'Astrophysique de Grenoble (IPAG) UMR 5274, Grenoble, F-38041, France
\and
Departamento de Fisica e Astronomia, Faculdade de Ci\^encias, Universidade do Porto, Portugal
             }

   \date{Received XX; accepted XX}

 
  \abstract
   { We define and put at the disposal of the community SOAP, \textit{Spot Oscillation And Planet}, a software tool that simulates the effect of stellar spots and plages on radial velocimetry and photometry.
    This paper describes the tool release and provides instructions for its use. 
    We present detailed tests with previous computations and real data to assess the code's performance and to validate its suitability. We characterize the variations of the radial velocity, line bisector, and photometric amplitude as a function of the main variables: projected stellar rotational velocity, filling factor of the spot, resolution of the spectrograph, linear limb-darkening coefficient, latitude of the spot, and inclination of the star.  Finally, we model the spot distributions on the active stars HD166435, TW Hya and HD189733 which reproduce the observations. We show that the software is remarkably fast, allowing several evolutions in its capabilities that could be performed to study the next challenges in the exoplanetary field connected with the stellar variability.
   }


   \maketitle
%

\section{Introduction}

Nowadays, more than 700 exoplanet candidates are reported in the literature, most of them detected \textit{via} the radial-velocimetry (RV) technique\footnote{Consult the up-to-date catalogue of the website \textit{exoplanet.eu}}. However, this efficient method is indirect as well as the photometric transit or astrometry techniques. One of the problems with this is the fact that  RV variations can in some cases be caused by other mechanisms that are not related to the presence of low-mass companions. Phenomena such as stellar pulsation, inhomogeneous convection, spots, or magnetic cycles can prevent us from finding planets, they might degrade the parameters estimation, or give us false candidates, if they produce a stable periodic signal (e.g. Queloz et al. 2001; Bonfils et al. 2007; Hu\'elamo et al. 2008). An essential work is then needed to understand all phenomena caused by stellar variability, and to characterize their impact on RV and photometry to be able to distinguish between these and real planetary signatures.

Particularly, dark spots and bright plages are present on the surface of dwarf stars from spectral types F to M, even in their low active phase (like the Sun, e.g. Meunier et al. 2010). Their appearance and disappearance on the stellar photosphere, combined with the stellar rotation, may lead to errors and uncertainties in the characterization of planets both in RV and photometry (e.g. Boisse et al. 2009). In a longer term, the change in the density of spots along the stellar magnetic cycle can also induce variations in RV and spectroscopic indices (Santos et al. 2010; Dumusque et al. 2011; Lovis et al. 2011).

Before the detection of the first planet (Mayor \& Queloz 1995), it was already noticed that the rotation of spots with the stellar surface alters the line profiles, modifying the line centroids without changing the true RV of the star. It was also thought that line bisector variations are also induced by modification of the convection pattern. In 1997, Saar \& Donahue developed simple models from a medium-strength Fe I line at about 6000\AA\ to examine the effect of starspots and convective inhomogeneities. The authors derived a relation between the semi-amplitude of the RV, the $v \sin i $ of the star and the fraction of the stellar disk covered by the spot. Using models of Ca I 6439~\AA\  and Fe I 6430 \AA\ lines, Hatzes (1999, 2002) reproduced the RV and the bisector line deformation induced by spotted line profile and derived similar relations. Desort et al. (2007) computed the visible spectra of various spectral type stars and simulated the weight of the spot in each wavelength, applying a black-body law to the stellar spectrum as a function of the spot temperature. To explain the variability observed in LkCa~19, Huerta et al. (2008) modeled the effect of starspots on the RV of young stars. To do so, they synthesized different spectra on a 100\AA\ bandpass near 6300\AA\ with the spot depicted as a zone with a lower temperature than the stellar surface, with lower intensity and different spectral features. In a different way, Lanza et al. (2010) derived a model to compute the RV variation from the photometric light curve that they applied on CoRoT-7 and HD189733 (Lanza et al. 2011). Aigrain et al. (2011) presented a similar model. 

Stabilized spectrographs give high-precision RV measurements from the cross-correlation function (CCF) of the spectrum with a numerical mask (Baranne et al. 1996, Pepe et al. 2002). The CCF is then fitted with a Gaussian to derive the stellar RV. The $FWHM$ and $contrast$ of the CCF were also recently found to be sensitive to short-term (Boisse et al. 2009; Queloz et al. 2009) and long-term (Santos et al. 2010; Lovis et al. 2011) activity. 
 
This paper presents SOAP, a tool offered to the community that enables simulating spots and plages on rotating stars and computes their impact on RV and photometric measurements. It also computes the main spectroscopic diagnostic to monitor stellar activity: the bisector span (BIS), calculated as defined by Queloz et al. (2001), and delivers a model of the CCF that is equivalent to the mean line of the spectrum. In particular, we highlight a novel method that enables a fast computation of the spot position and geometry.

The next section describes the software, giving the platform, the language, a short explanation of the principle of the code, a complete description of the input and output parameters, and finally, some details of the computing. We expose in a third section different tests to compare the results with other simulations and show the relation between the variations in RV, BIS and photometry with the available parameters of the code. We then apply SOAP to the data of HD166435 (Queloz et al. 2001), TW Hya (Hu\'elamo et al. 2008; Donati et al. 2011) and HD189733 (Boisse et al. 2009). The last section discusses possible applications and future improvements that can be done with SOAP.


\section{Software description}

	\subsection{Platform and language}
The SOAP software is available for use at {\texttt{http://www.astro.up.pt/soap} along with a brief manual and some example data.
When using SOAP in publication, it is appropriate to cite this paper. 

SOAP is written with both C and python, for the core and high-level functions, respectively. An input parameter file  \texttt{driver.cfg} has to be filled in by the user and uploaded on the webpage. After running the program, an output file, \texttt{output.dat}, and an output directory, \texttt{CCFs}, are returned.

	\subsection{Principle}
SOAP first computes the non-spotted emission of the star in photometry and in RV. To do that, the visible stellar disk is centered on a grid of several tens of thousands of resolution elements. The typical line of the emerging spectrum of the star is modeled by a Gaussian for each grid cell that is equivalent to the spectrum's CCF. Each Gaussian (in a given stellar position) is Doppler-shifted according to the projected rotational velocity and weighted by a linear limb-darkening law. 

SOAP calculates the position of the surface inhomogeneities defined by their latitudes, longitudes, and sizes. The cells inside the spots are modeled by the same Gaussian as for the stellar disk, but are weighted by their brightness (0 for a dark spot, higher than 1 for a bright plage). The code then removes (for dark spots) or adds (for bright spots or plages) the CCF and flux of the inhomogeneities to those of the non-spotted star. Finally, SOAP delivers the integrated spectral line, the flux, the RV, and the BIS as a function of the stellar rotational phase.

	\subsection{Input parameters}
	
	The input parameters are assigned by the user in the file \texttt{driver.cfg}. They are displayed in Table~\ref{input} and described in the following. 

\begin{description}
	\item[Resolution of the simulation:]
	
	The parameter $grid$ determines the number of cells to resolve the stellar disk on a squared box of $grid$$\times$$grid$ resolution elements. The position of the spot at each phase is settled thanks to the determination of the location of its circumference (cf. Sect.3.2.2). The spot circumference is determined with  $nrho$ number of resolution elements (cf. Fig~\ref{nrho}). 
	 We find that there are no significant changes in the results for values beyond $grid$=300 and $nrho$=20, and conclude that this option combines fast computation and sufficient resolution.\\

   \begin{figure}
   \centering
   \includegraphics[width=8.5cm]{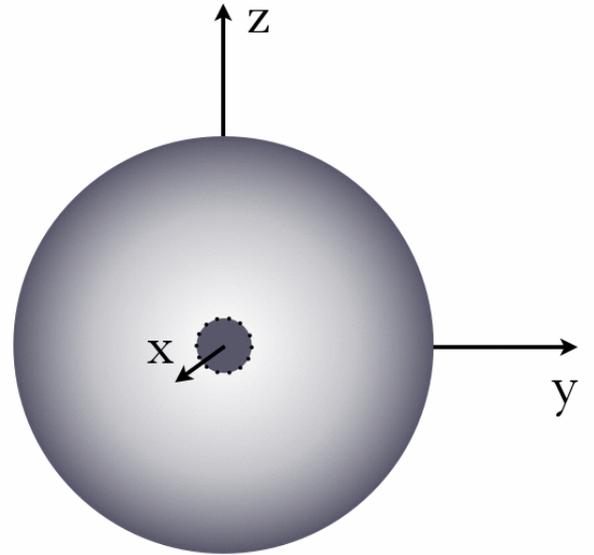}
      \caption{ In three dimensions, the spot is first located in front of the observer (the $x$-axis is aligned with the line of sight). $nrho$ number of points are equally distributed along the circumference of the spot (small black dots). 
             }
         \label{nrho}
   \end{figure}

	\item[Stellar parameters:]
	The user can change the following stellar parameters: linear limb-darkening coefficient $limb$, stellar radius $radius$ expressed in Sun radius, rotation period $Prot$ in day, stellar inclination angle with the line of sight $I$, mean star's velocity $gamma$ in kms$^{-1}$, and the initial phase for the computation of the simulation, $psi$.  We note that SOAP fixes the reference of 0$^{\circ}$ in longitude as the position of the first spot in front of the line of sight at phase equals to 0.\\
	
	\item[Spot parameters:]
	The code allows including up to four spots, and choosing their longitude $long$ and latitude $lat$ in degrees. Their radius is defined by $size$ and expressed as a linear projected value of the stellar radius.  We remark that for a filling factor of less than 10\%, a radius expressed as a linear projected value or as an arc value along the surface is completely identical. The brightness of the spot, defined by $bright$, corresponds to a weight to the stellar flux, and could take all positive values. Set to 0$\leqslant bright<1$, it will model a darker spot than the stellar photosphere and, with a value higher than 1, it will simulate plages with a higher brightness than the star. This weight is applied to the CCF that modeled the emitted spectrum. We note that the same Gaussian is used for the photosphere and the inhomogeneities. 	\\

	\item[CCF parameters:]
	The parameters of the Gaussian can also be chosen to account for the resolution and accuracy of the spectrograph, as well as the way the observed CCF was computed to be as close as possible to real data. The code asks for the width in kms$^{-1}$, $sigma0$, of the CCF of a non-rotating star (or with rotational velocity too low to be resolved) with the same spectral type. We recall that it is related to the $FWHM$ by $\sigma = \frac{FWHM}{2\sqrt{2\ln2}}$ in kms$^{-1}$. The two other parameters, $window$ and the $step$ in kms$^{-1}$, give the characteristics to calculate the CCF  function. For low-rotator stars and high-resolution spectrographs, typical values are $window$=20kms$^{-1}$ and $step$=0.5kms$^{-1}$ (see Fig.~\ref{CCF}). Higher values should be used for $window$ for young and/or fast-rotators stars. 

\end{description}
	 
   \begin{figure}
   \centering
   \includegraphics[width=8.5cm]{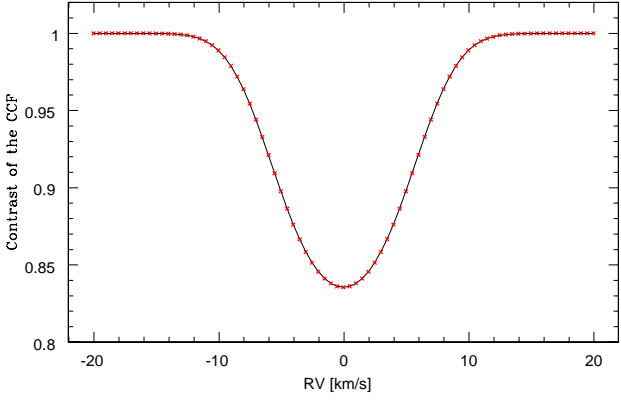}
      \caption{ Gaussian that modeled the CCF. The contrast of the CCF corresponds to the amplitude of the Gaussian. The window in RV to compute the CCF is the $x$-axis, here the $window$ parameter is equal to $20$~kms$^{-1}$. The red points illustrate the RV where the CCF values are calculated. The span between two points is the $step$ parameter, here equal to $0.5$~kms$^{-1}$.
             }
         \label{CCF}
   \end{figure}

\begin{table}
  \centering 
  \caption{Input parameters}
  \label{input}
\begin{tabular}{lll}
\hline
\hline
Parameter   &  Typical value        &  Comments       \\
\hline
\hline
  grid      & 300      &   grid$\times$grid resolution elements  \\
   nrho     &  20     &          \\
\hline

limb   &   0.5 to 0.9       & linear limb-darkening coef.     \\
radius [R$_{\odot}$] &             &    \\
Prot [day]   &   &  \\
I [$^{\circ}$]       & 0 to 90   &   stellar inclination  \\
gamma  [kms$^{-1}$] & 0. & mean star's velocity \\
psi & 0. & \\
\hline

long [$^{\circ}$]  & 0. & \\
lat [$^{\circ}$]  & -90 to +90 & SOAP simulates up to 4 spots\\
size  [R$_{\star}$] & &   spot size in stellar radius\\
bright & 0 to $>$1 &  dark spot=0 ; plages$>$1\\
\hline

prof0 & 0.4 & \\
sigma0 [kms$^{-1}$] & 2 to 5 & spectrograph resolution \\
window [kms$^{-1}$] & 20 & window calculating CCF function\\
step [kms$^{-1}$] & 0.1 & \\
\hline            
\hline%
\end{tabular}
\end{table}

	\subsection{Output parameters}
	\label{output}
	The format of the output parameters is determined in the file \texttt{driver.cfg}. 
SOAP gives the values of the flux, RV and, BIS, as a function of the phase step and writes the result in the output file $file\_out$.  By default $file\_out$ is named 'output.dat'. It gives a file with four columns: phase, flux, RV, and BIS. These values can be directly plotted by SOAP as a function of the stellar rotational phase if screen output is selected (Fig.~\ref{FigSOAP}). 
	
 The phase step can be chosen in two ways. It can be a constant step of a fraction of the phase, $ph\_step$ between 0. to 1. If this last is set equal to 0, the code will read phases in a file determined as an input parameter $ph\_in$.
	
	The computed CCFs can also be stored in the directory named CCFs if the $ccf\_out$ is set to 1 (no output = 0). All input parameters are archived in the header of the FITS files generated. Each contains a one-dimension FITS file of the flux as a function of the RV on a $window$ and with a $step$ determined by the input parameters of the CCF.

   \begin{figure}
   \centering
   \includegraphics[width=8.5cm]{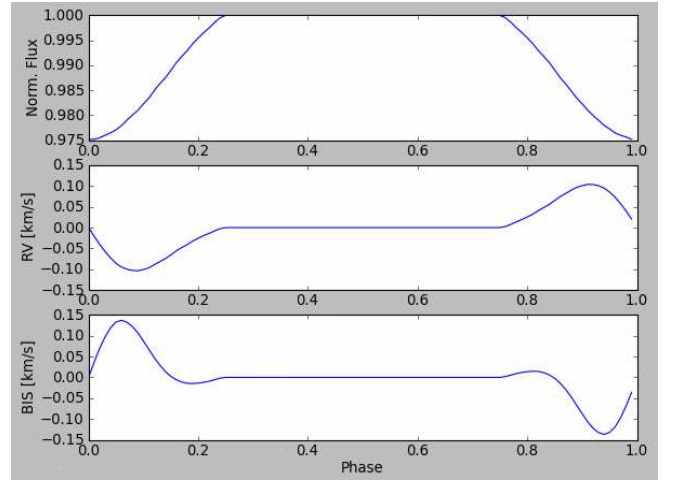}
      \caption{ Figure generated by SOAP (optional) showing the photometry, the RV and the BIS (from \textit{top} to \textit{bottom panel}) as a function of the stellar rotational phase. As a default parameter, the first spot is in front of the line of sight at phase 0.          }
         \label{FigSOAP}
   \end{figure}

	\subsection{Detailed numerical computing}
		\subsubsection*{CCF and flux of the non-spotted star}
		The star is computed as a disk with a radius normalized to $1$. This disk is centered on a grid of $grid\times grid$ cells on a $y$-$z$ plane. The cells take their $y$ and $z$ values between $-1$ and $1$. Each grid cell ($p_{y}, p_{z}$) contains a flux value and a CCF. The flux value depends on the linear limb-darkening law. The CCF is modeled by a Gaussian with width and amplitude given by the input parameters, Doppler-shifted according to the projected rotational velocity and weighted by the linear limb-darkening law. The projected rotational velocity is calculated from the rotational period, stellar radius and inclination of the spin axis with the line of sight, given in the input parameters.
		
		The code first sums the contribution of all grid cells of the non-spotted star to generate reference CCF and flux value.

		\subsubsection*{Position and contribution of the inhomogeneities ("spots")} 
	      The spot is defined in the simulation as a disk on the stellar surface. When the spot rotates with the stellar photosphere, the shape of the spot seen along the line of sight changes and becomes ellipsoidal. The code is able to compute this so fast (see Sect.~\ref{performance})  because it does not calculate the projected geometry of the spot.Ê\\
	     
	     The grid is on the $y$-$z$ plane and the $x$-axis is along the line of sight. If the star is seen edge-on, it rotates around the z-axis. We remember that the stellar angle in the plane of the sky has no impact on the result and is completely omitted. The grid cells take values between $-1$ and $1$. These maximum values correspond to the stellar radius. \\

   \begin{figure*}
   \centering
   \includegraphics[width=4.5cm]{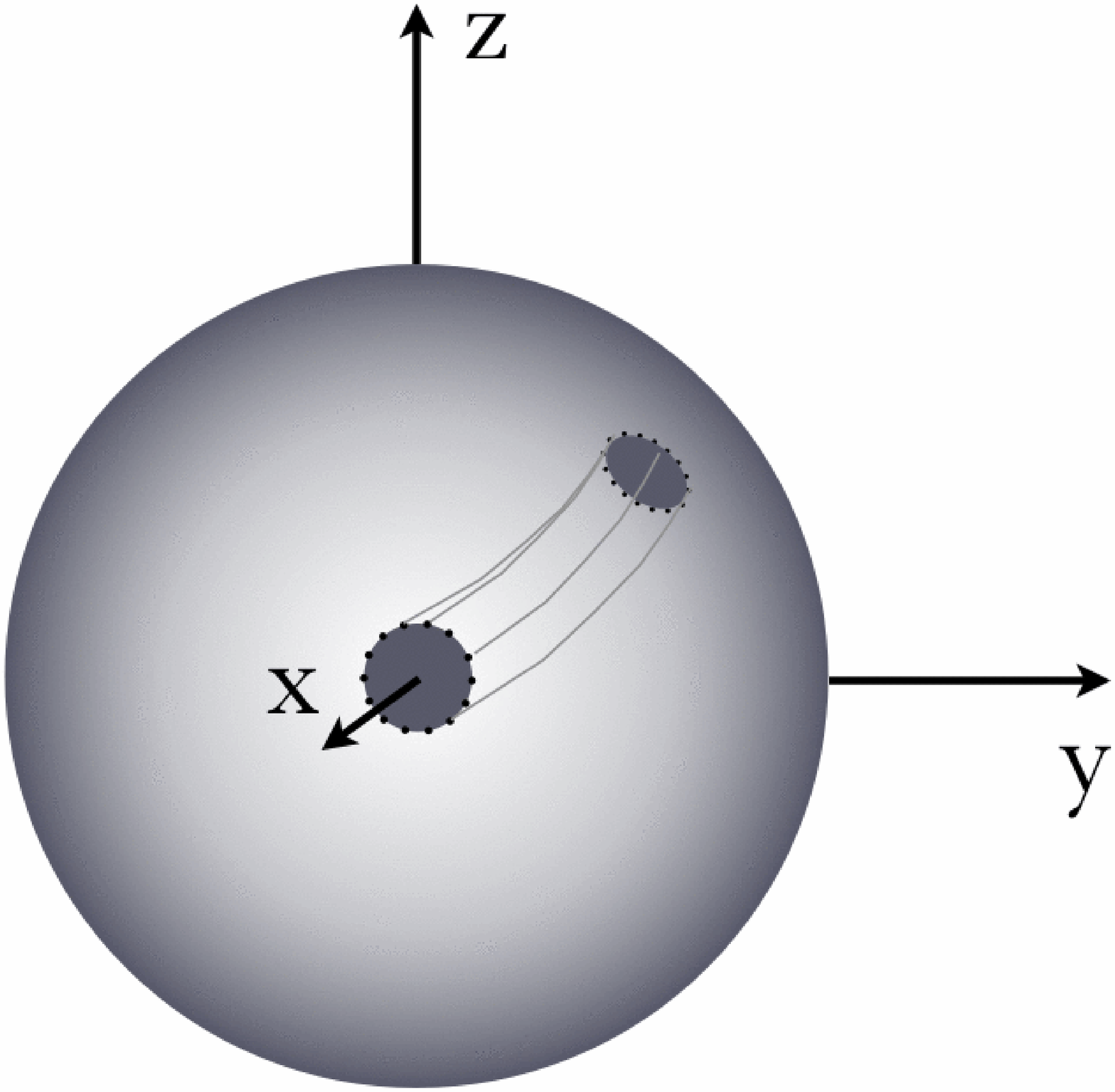}
     \includegraphics[width=4.5cm]{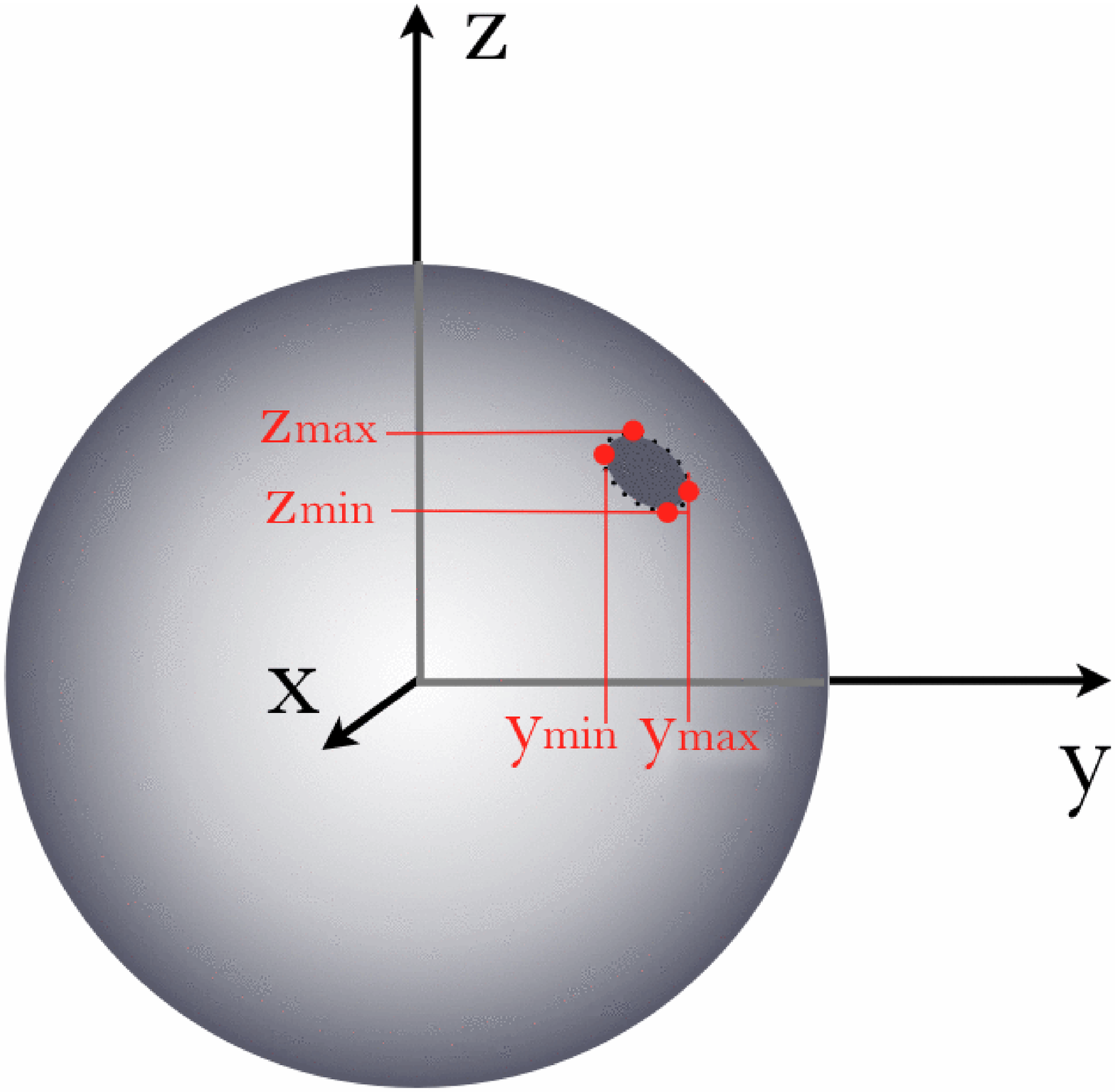}
   \includegraphics[width=4.5cm]{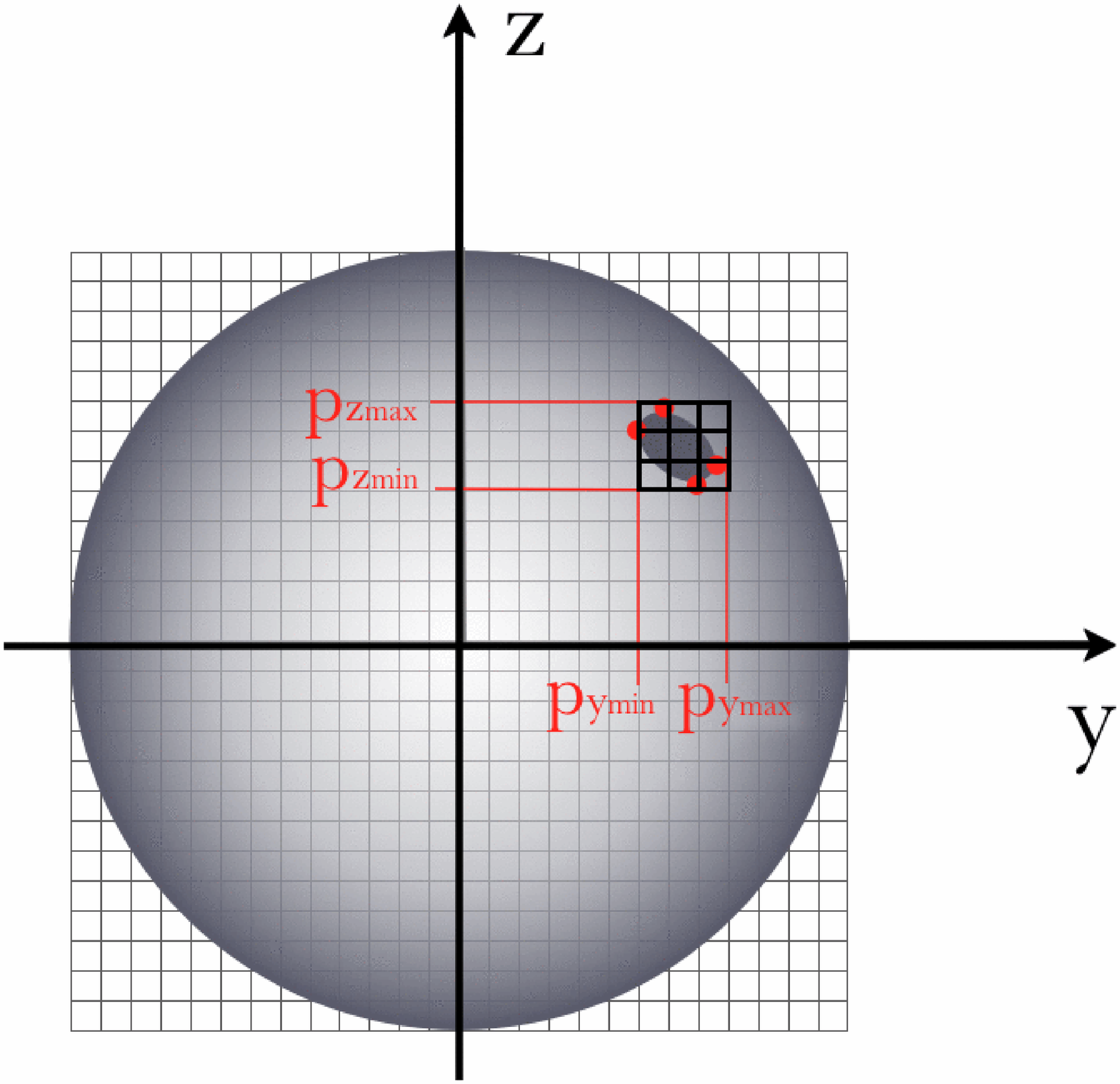}
   \includegraphics[width=4.5cm]{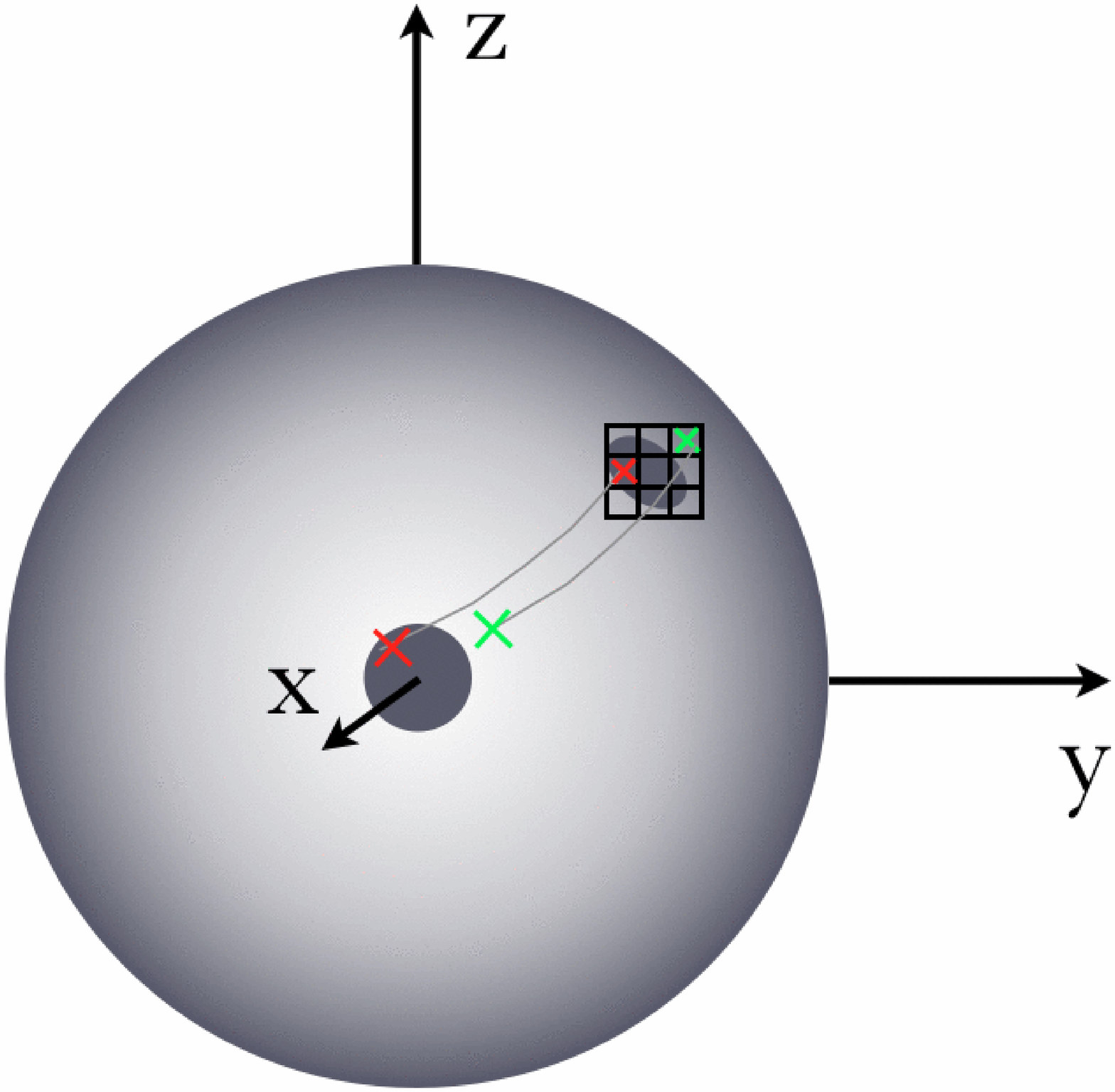}
    \caption{  \textit{From left to right:} 1) According to the stellar inclination and to the spot's latitude and longitude, the $xyz$ position of the $nrho$ points on the spot's circumference are calculated in 3-D. 2) The maximum and minimum $y$ and $z$ determine a smaller area where the spot is. 3) These \{$yz$\} positions (in 3D) are projected on the $y$-$z$ plane, which gives ($p_{y}$,$p_{z}$) positions. 4) From these values, a smaller grid is defined. Each cell of this grid is scanned. A reverse to that performed in step 1 rotation is performed to establish if the cell is inside the spot or not.
             }
         \label{calculation}
   \end{figure*}

		         For each phase, the code first determines the area of the grid where the inhomogeneities are located. To do that, the spot is initialized in front of the line of sight, on the equator of a star seen edge-on. From there, the spot has a well known geometry of a circle centered on the $x$-axis and a radius $size$ (in stellar radius) given in the input parameters.
         
         For each $nrho$, the number of points on the circumference of the spot, the code calculates the position according to the stellar inclination, spot latitude and longitude (see step 1 in Fig.~\ref{calculation}). For that a three-rotations matrix is applied to the $nrho$ points: a rotation of the angle of the stellar inclination $i$ about the y-axis $R_{y}(\varphi)$, a rotation of the spot longitude $long$ about the z-axis $R_{z}(\theta)$, and a rotation of the spot latitude $lat$ about the y-axis $R_{y}(\psi)$:	
\begin{equation}
R_{y}(\varphi)\times R_{z}(\theta)\times R_{y}(\psi)=  \\
\left|\begin{array}{ccc} a & -\sin\theta \sin\varphi & b \\ \sin\theta \cos\psi & \cos\theta & \sin\theta \sin\psi \\  c & \sin\varphi \sin\theta & d  \end{array}\right|
\label{rotation}
\end{equation}         
with
\begin{eqnarray*}
\lefteqn{ a = \cos\varphi \cos\theta \cos\psi-\sin\psi\sin\varphi} \\
\lefteqn{ b =  \cos\varphi \cos\theta \cos\psi+\sin\varphi \cos\psi} \\
\lefteqn{ c = -\sin\varphi \cos\theta \cos\psi-\cos\varphi \sin\psi} \\
\lefteqn{ d = \cos\varphi\cos\psi-\sin\varphi\cos\theta\sin\psi.} 
\end{eqnarray*}
	
	 We note that the conventions give $\varphi=90-i$ (in degree) and $\psi=-lat$ and at that point, the $xyz$ values are given as a fraction of the stellar radius.
	
	The $xyz$ positions of the $nrho$ points are then rotated according to the rotational phase, with an angle $\phi$=360$\times$$ph$, $ph$ is a number between 0 and 1 (see Section\ref{output}). The rotation is performed along the unit vector aligned with the stellar inclination $\vec{u}$=($\cos$($i$), 0, $\sin$($i$)). The rotation matrix $R_{\vec{u}}(\phi)$ is
\begin{equation}
R_{\vec{u}}(\phi)=\left|\begin{array}{ccc}\alpha \mathrm{u_{x}^{2}}+\cos\phi & \alpha  \mathrm{u_{x}u_{y}}-\beta  \mathrm{u_{z}} & \alpha  \mathrm{u_{x}u_{z}}+\beta  \mathrm{u_{y}} \\\alpha  \mathrm{u_{y}u_{x}}+\beta  \mathrm{u_{z}} & \alpha  \mathrm{u_{y}^{2}}+\cos\phi & \alpha  \mathrm{u_{y}u_{z}}-\beta  \mathrm{u_{x}} \\ \alpha  \mathrm{u_{z}u_{x}}-\beta  \mathrm{u_{y}} & \alpha  \mathrm{u_{z}u_{y}}+\beta  \mathrm{u_{x}} & \alpha  \mathrm{u_{z}^{2}}+\cos\phi \end{array}\right|
\end{equation}
with
\begin{eqnarray*}
\lefteqn{\alpha = (1-\cos\phi)}\\
\lefteqn{\beta = \sin\phi}.
\end{eqnarray*}	
	
	From these $nrho$ $xyz$ positions, the code defines the visibility of the spot. If it is visible, it defines a smaller area where the spot is according to the minimum and maximum values taken by $y$ and $z$ (see step 2 in Fig.~\ref{calculation}). When there are both visible and invisible points, i.e. the spot is on the edge, the minima/maxima are over-/underestimated if the spot is on one of the axis ($y$ or $z$). This is solved following the idea that if the spot is on the bottom z-axis ($z<0$), then the minimum z value is equal to $-1$. 
		
	Then, the maximal/minimal $y$-$z$ values are transposed onto the grid ($p_{y}, p_{z}$), which depends on the sampling given by the $grid$ parameter (see step 3 in Fig.~\ref{calculation}). We note that the code never memorizes the position of the spot in the grid coordinate ($p_{y}, p_{z}$). \\
	
	Then the code calculates the impact of the spot on the stellar photometry and velocity. The code scans the smaller grid and checks whether each center of the cell is located inside or outside the spot. Because we do not know the projected geometry of the spot in its actual position, we perform an inverse rotation of the matrix given in Eq.\ref{rotation} to replace the grid point where it would be in the initial configuration, an equatorial spot on an edge-on star, where the spot has the well-known geometry of a circle centered on the x-axis (see step 4 in Fig.~\ref{calculation}). 	
	If the grid-point ($p_{y}, p_{z}$) is inside the spot, a Gaussian is attributed to this point with a velocity according to the projected rotational velocity, and weighted by the linear limb-darkening law and the intensity of the spot defined by its brightness $(1-bright)$. For this last point, we note that the computed CCF for the spots	is to represent what is removed (for dark spots) or added (for bright spots) to the unspotted stellar CCF.
	The code then generates a summed CCF and a flux value for the spots at each wanted rotational phase. For dark spots, the CCF and flux values correspond to the stellar intensity that is hidden by the spots.

		\subsubsection*{Final parameters}
		\label{final}
		The code then subtracts for each phase the spots CCF from the non-spotted stellar CCF, and the spots flux value from the non-spotted stellar flux. SOAP returns a CCF and a flux value for each queried phase.
		
		The flux value corresponds to the photometry, which is normalized to the maximum value along the rotational phase.
		
		Each CCF is normalized to the maximum value of each CCF. Each CCF is fitted by a Gaussian to return RV values for each phase. The BIS values are calculated as described in Queloz et al. (2001).

\section{Tests and comparison with data}

	\subsection{Performance}
	\label{performance}
	The computation of the code is sufficiently fast to allow many simulations and adjusting the data. For a MacBook Pro with a 2.33 GHz Intel Core Duo, calculating 10,000 phases of a star with one always visible spot takes less than 40~s. To simulate 10,000 phases of a star with four spots, the calculation and the output files are ready in less than 90~s. We emphasize that in the simulations reported below (Sect.4.3), a sufficient number of 100 phases per rotational period were computed. The code is therefore perfectly suitable for solving inverse problems.

	\subsection{Looking at different parameters}
	In Boisse et al. (2011), we used SOAP to derive the periodicity properties of the RV variations induced by spots, as well as to derive the relations between the RV, the photometry and the CCF parameters (BIS, FWHM and contrast). For the simulations, we chose a G0V star with a radius of 1.1 R$_{\odot}$ and a $v \sin i$=5.7kms$^{-1}$, a linear limb darkening of 0.6, and a spectrograph resolution of 110~000. The spot was considered as a dark surface without any emission of light, and a size of 1\% of the visible stellar surface. We refer to this paper for the computation of the RV variations induced by one or two spots with different stellar inclinations and spot latitudes and longitudes (cf. their Figures 1 and 11), and also for the corresponding relations between the RV and the bisector span (cf. their Figures 6 and 13). These authors also showed the relations between the FWHM and the photometry for one and two spots (their Figures 8 and 14). For one specific case, the relation between RV, BIS, FWHM, contrast, and photometry is shown in their Figures 1 and 7. On the webpage, one can find computed values (output files) for specific cases so that the user can check the proper calculation of the code.
	
	In Fig.~\ref{FigSOAP}, we show the flux, RV, and BIS as a function of the stellar rotational phase for one equatorial spot of 1\% on the surface of an edge-on star with equivalent parameters that are given in the previous paragraph. \\

   \begin{figure}[t]
   \centering
  \includegraphics[width=8.5cm]{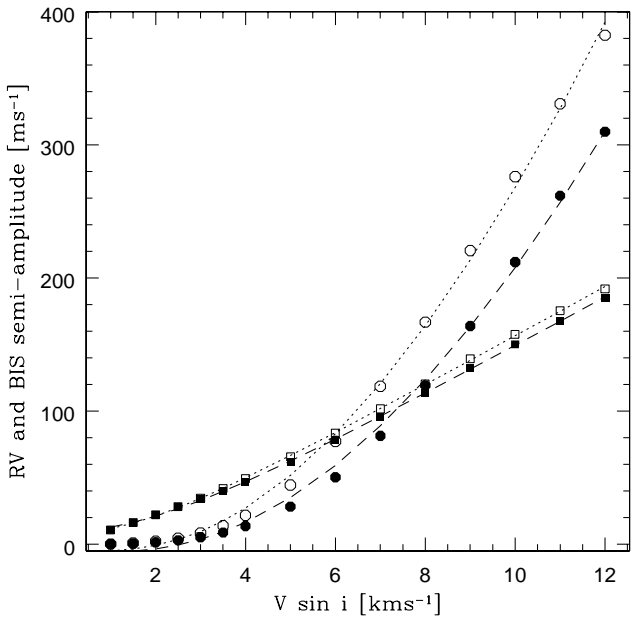}
      \caption{ Semi-amplitude of the RV (squares) and BIS (circles) variations as a function of $v\sin i$. The empty points are for a resolution power of the spectrograph of R$=75,000$ and the filled ones are for R$=40,000$. All other parameters are fixed (cf. text). The lines are the best chi-square fits from equations \ref{depRV} and \ref{depBIS}.
             }
         \label{FigVsini}
   \end{figure}

	To check the reliability of the code with previous simulations, we sought to reproduce the laws derived first by Saar \& Donahue (1997), and by Hatzes (1999, 2002) and Desort et al. (2007) (cf. introduction). We simulated a spotted star as close as possible to the star of the Saar \& Donahue (1997) simulation. We generated a 1\% dark spot on the equator of a G2V edge-on star. In the following analysis of this section, all parameters are fixed on these values, except those studied in each paragraph. We plot in Fig.~\ref{FigVsini} the semi-amplitude of the RV (squares) and the BIS (circles) as a function of the $v \sin i$.  The semi-amplitude is defined in this paper as half of the maximum deviation, hence $\frac{\Delta \mathrm{RV}}{2}$, $\frac{\Delta \mathrm{BIS}}{2}$, and later $\frac{\Delta \mathrm{flux}}{2}$ for the photometry. 
	We considered observations made with a spectrograph with a resolution power of about R$=75,000$ (empty points) and R$=40,000$ (filled ones). We recall that the resolution power is coded in SOAP as $sigma_{0}$, the width of the CCF of a non-rotating star or with a rotational velocity too low to be resolved. The $sigma_{0}$ values of G0 star are derived from the Eq.B.2 and B.3 of Boisse et al. (2010) for a spectrograph resolution of 75,000 and 40,000. In contrast with previous results, we observe that the RV depends not only on the $v \sin i$ but also slightly on the resolution of the spectrograph. As shown by Desort et al. (2007), the BIS variations also depend on $v \sin i$ and on the resolution of the spectrograph. 
	This dependence on the instrumental resolution is explained because the RV and the BIS are derived from the fit of a Gaussian on the CCF (cf. Section~\ref{final}) and the deformation of the CCF induced by a spot of a given size depending on the instrumental resolution.

	We derived the relation between the semi-amplitude of the variation of the RV, BIS, and filling factor of the spot $f_{r}$, plotted in Fig.~\ref{FigFr}. 
	
   \begin{figure}
   \centering
   \includegraphics[width=9cm]{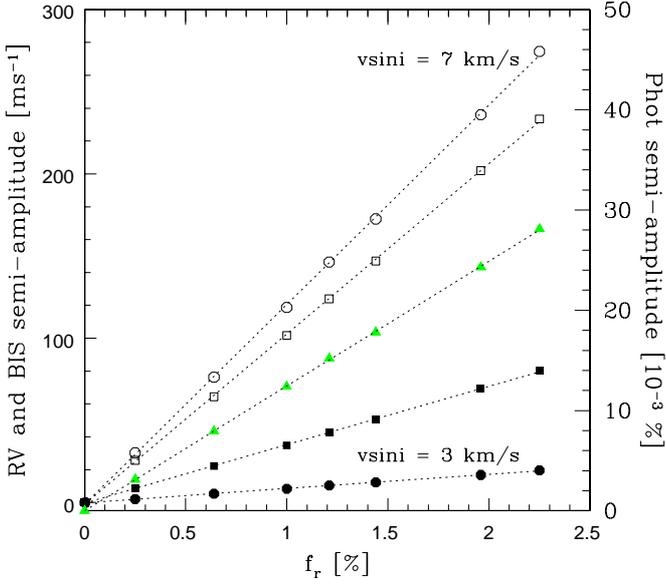}
      \caption{ Semi-amplitude of RV (square), BIS (circle), and photometry (green triangle) as a function of the filling factor of the spot $f_{r}$. Two different $v\sin i$ at 3 and 7 kms$^{-1}$ (filled and empty markers) are drawn. The photometry does not depend on $v \sin i$. All other parameters are fixed (cf. text). The dashed lines correspond to the best linear fits.
             }
         \label{FigFr}
   \end{figure}

	We obtained the following laws:
	
\begin{equation}
\frac{\Delta \mathrm{RV}}{2} \approx 20.67\, f_{r} \, (\sqrt{v\sin i^{2} + sigma_{0}^{2}} - sigma_{0})^{0.98} 
\label{depRV}
\end{equation}	
	
\begin{equation}
\frac{\Delta \mathrm{BIS}}{2} \approx 9.6 \, f_{r} \, (\sqrt{v\sin i^{2} + sigma_{0}^{2}} - sigma_{0})^{1.65}. 
\label{depBIS}
\end{equation}	 

	These relations give an equivalent range of the amplitudes to those derived by Desort et al. (2007). The differences emerge because we did not simulate a temperature for the spot. We remark that Desort et al.'s results are close to those of Saar \& Donahue (1997) and Hatzes (1999, 2002) at low $v\sin i$ but then they depart strongly. As Desort et al. (2007) noted, this is because Saar \& Donahue (1997) and Hatzes (1999, 2002) modeled the effect in only one single line that has a varying sensibility to the impact of spots. Hatzes (1999) also noticed that. As already mentioned by Desort et al. (2007), these estimates, derived by us in Eqs.~\ref{depRV} and~\ref{depBIS} or by the previously quoted authors, are indicative and we discourage quantitative conclusions derived from the blind use of one of these equations. 

	In Fig.~\ref{FigVsini}, we observe that when $v\sin i$ is lower than a certain value, depending on the resolution of the spectrograph, the bisector does not vary, and thus cannot be used as a diagnostic of stellar activity. This phenomenon was already pointed out by Santos et al. (2003) and was already observed for active stars with a very low rotational velocity (e.g. Bonfils et al. 2007).

   \begin{figure}
   \centering
   \includegraphics[width=8.5cm]{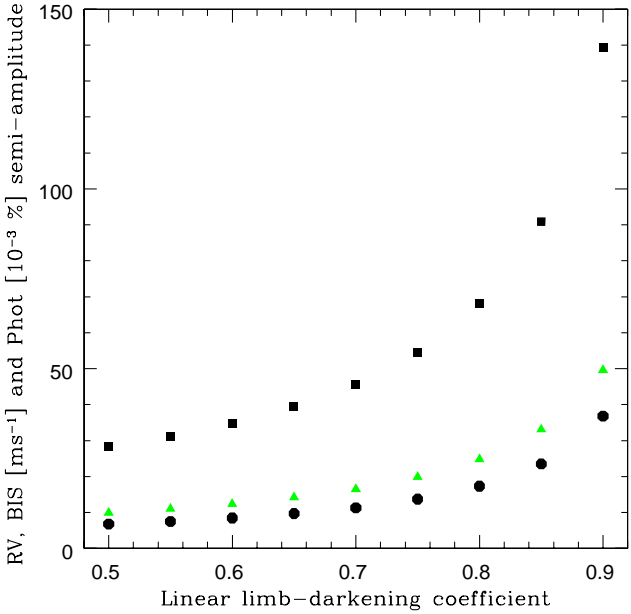}
      \caption{ Semi-amplitude of RV (square), BIS (circle), and photometry (triangle) as a function of the linear limb-darkening coefficient. All  other parameters are fixed.
             }
         \label{FigLimb}
   \end{figure}

	We also aimed to characterize the impact of the linear limb-darkening coefficient. Following Claret (2004), this coefficient varies roughly from 0.5 to 0.9 (in V band) for dwarf stars with an effective temperature between 6500~K and 3800~K. 
	Fig.~\ref{FigLimb} illustrates that for the same spot, a higher linear limb-darkening coefficient induces an increased impact in RV, BIS, and photometry. This is mainly explained by a total lower brightness of the spotless star.
		
	Figs.~\ref{FigLat}-\ref{FigI} show the variations in the amplitude of the RV, BIS, and photometry as a function of stellar inclination and spot latitude angles. We remark that the variation laws are close to a  $\cos^{2}$, as drawn with dashed lines in the figures, as expected because of  the effects of projection. 
	
	We emphasize that in our simple model, most of the variables are degenerate (like the size and the brightness).   
	Nevertheless, some parameters can be constrained by observations that allow to model real data.

   \begin{figure}
   \centering
   \includegraphics[width=8.5cm]{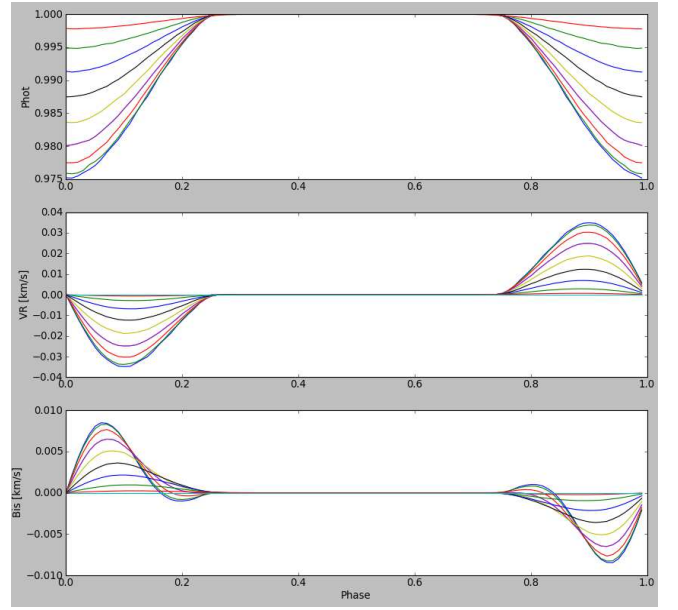}
      \caption{ RV, BIS, and photometry as a function of the stellar rotational phase. The different colors illustrate the change in latitude of the spot. The maximum amplitude variation of the three parameters are given for an equatorial spot. No variation is detected when the spot is on the pole. We note that beyond a certain value of $lat$ ($\approx$~50$^{\circ}$), the shape of the BIS variation changes.
             }
         \label{FigLat}
   \end{figure}

   \begin{figure}
   \centering
   \includegraphics[width=9cm]{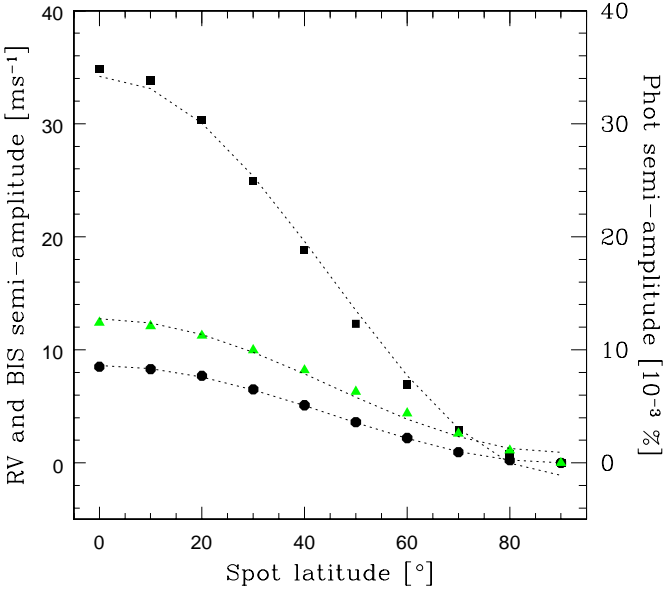}
      \caption{ Semi-amplitude of RV (square), BIS (circle), and photometry (triangle) as a function of the spot latitude. All the other parameters are fixed.  The dashed lines correspond to a $\cos(i)^{2}$ fit.
             }
         \label{FigLat2}
   \end{figure}

   \begin{figure}
   \centering
   \includegraphics[width=9cm]{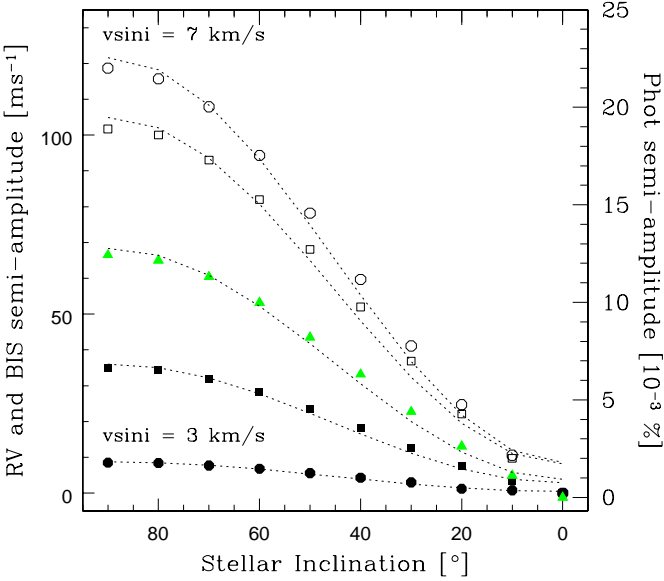}
      \caption{ Semi-amplitude of RV (square), BIS (circle), and photometry (green triangle) as a function of the inclination of the star with the line of sight $i$. Two different $v\sin i$ at 3 and 7 kms$^{-1}$ (filled and empty markers) are drawn. The photometry does not depend on $v \sin i$. All other parameters are fixed (cf. text). The dashed lines correspond to a $\cos(i)^{2}$ fit.
             }
         \label{FigI}
   \end{figure}

	\subsection{Simulations of published spotted stars}
	Several publications reported stellar RV variations that are caused by magnetic activity and not to a gravitationally bound companion. These cases could be reproduced by SOAP, which sheds light onto the parameters of the spotted stars. 
	
		\subsubsection*{HD166435}	
	
	Queloz et al. (2001) published RV variations with a period of 3.7987 days of HD166435 observed during two years. The star was found to be photometrically and magnetically variable with the same period. This and the long-term phase instability of the RV signal, are signatures of the evolution of a spot in a stable active region. 
	We simulated the HD166435 variability with one spot on the stellar surface. This G0V star was observed with ELODIE, a spectrograph with a resolution power of about 42,000. Queloz et al. (1998) published for this spectrograph the relation between the stellar ($B-V$) and $sigma_{0}$, the width of the CCF of a non-rotating star or with a rotational velocity too low to be resolved. We used their equation (2) and the HD166435 ($B-V$)=0.633 value to derive a $sigma$0 equal to 4.62~kms$^{-1}$.
	We chose a linear limb-darkening coefficient of 0.6. Queloz et al. (2001) derivedthe stellar $v \sin i$=7.6$\pm$0.5kms$^{-1}$ from the width of the CCF. With a radius of 0.99R$_{\odot}$ (Takeda 2007), we can then derive a stellar inclination of 35$^{\circ}$ with the line of sight. The brightness and the size parameters are almost completely degenerate for small spots. Therefore, we decided to fix $bright$ to 0. and to adjust the size of the spot. Because only smooth variations of the RV, BIS, and flux (without plateau) are observed, this suggests that the spot is always visible to the observer. Using the simulations, smooth variations of the parameters are observed for a spot with a latitude greater than 45$^{\circ}$ at a stellar inclination of 35$^{\circ}$. Finally, we varied the latitude and size of the spot to be as close as possible to the observations. To obtain a good ratio between the RV and BIS semi-amplitude, we need to have a spot latitude close to 68$^{\circ}$. The nearest semi-amplitude values to the observations are then obtained for a spot size of 2.02\% of the stellar disk. The related normalized flux varies by 2.9\%, which corresponds for a relative magnitude of 6.85, to 0.033mag, which is in the lower range of the observed 0.035 to 0.05mag variation. If we fit the RV variations with a Keplerian model, an eccentricity of 0.22 is found, equivalent to the 0.2 value reported by Queloz et al. (2001). We then agree with the conclusion of Queloz et al. (2001) that a simple one-spot model can well reproduce the observed variability.
	In Fig.~\ref{FigHD166435}, we show the result of our simulation with the input parameters given in Table~\ref{queloz}.
	We tested if a two-spot scenario at opposite longitudes could also explain the data. The period would then be equal to 7.5974d. With a radius of 0.99R$_{\odot}$, the  $v \sin i$ should be overestimated. We chose to put the star equator-on, which corresponds to a $v \sin i$ of 6.6~kms$^{-1}$. We fixed the brightness of the spot to 0. We ran the simulations and found that no solution leads to the measured values of the semi-amplitude of the RV and the BIS (and to a good ratio between the RV and BIS semi-amplitude variations). We conclude that the two-spot scenario is not appropriate.

   \begin{figure}
   \centering
  \includegraphics[width=8cm]{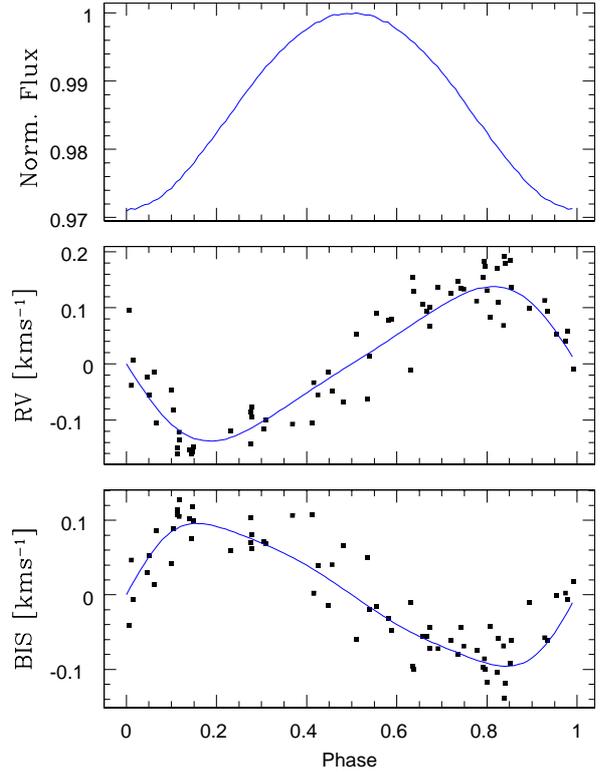}
      \caption{SOAP simulation of a star with the same characteristics as HD166435 observed with ELODIE with a spot of 2.02\% of the stellar surface to reproduce the RV, BIS, and flux variations observed by Queloz et al. (2001). 
             }
         \label{FigHD166435}
   \end{figure}

\begin{table}
  \centering 
  \caption{Input parameters for the simulation of HD166435 }
  \label{queloz}
\begin{tabular}{lll}
\hline
\hline
Parameter   &  Value        &       Ref.   \\
\hline
\hline
  grid      & 300      &     \\
   nrho     &  20     &          \\
\hline

limb   &   0.6       &      G0V star \\
radius [R$_{\odot}$] &  0.99           &   Takeda 2007 \\
Prot [day]   &  3.7987 & Queloz et al. 2001 \\
i [$^{\circ}$]       &  35   &   calculated (see text)  \\
gamma  [kms$^{-1}$] & 0. & \\
psi & 0. & \\
\hline

lat [$^{\circ}$]  & 68 & \\ 
bright & 0. & fixed \\
size  [R$_{\star}$] & 0.142 \\
\hline

sigma0 [kms$^{-1}$] & 4.62 & ELODIE resolution \\
window [kms$^{-1}$] & 20 & \\
step [kms$^{-1}$] & 0.1 & \\
\hline            
\hline
\end{tabular}
\end{table}
	
	\subsubsection*{TW Hya}	
	
	The classical T Tauri star, TW Hya, was first announced to host a planet (Setiawan et al. 2008) before infrared RV measurements showed that a cold spot covering $\sim$7\% of the stellar surface and located at a latitude of 54$^\circ$ better explained the observations (Hu\'elamo et al. 2008). Indeed, the infrared RV variations presented a lower amplitude than the optical ones, in agreement with the spot scenario. The contrast between the spot and the photosphere is weaker in the near-infrared than in optical, inducing weaker RV variations. The spot model was derived with SOAP using the optical observations. All parameters needed for the simulations are given in the Section 2 of Hu\'elamo et al. (2008). These authors reproduced an observed semi-amplitude of about 150ms$^{-1}$. The model predicts a photometric variation of 4\%, much lower than the observed 20\%, which prompted the authors, among other indices, to conclude that there is a dark spot combined with a bright one due to accretion. 
	  Recently, Donati et al. (2011) observed TW Hya with the ESPaDOnS spectropolarimeter at the CFHT. These authors confirmed that the RV fluctuations should come from a high-latitude photospheric cool spot overlapping with the main magnetic pole where most of the accretion is concentrated. The authors observed peak-to-peak RV variations of  $\approx$50ms$^{-1}$ that they assumed to originate from a spot at high-latitude with a size of 2\% of the stellar surface. We used SOAP to simulate their data with the input parameters given in Table~\ref{donati}. As observed by Setiawan et al. (2008) and Hu\'elamo et al. (2008), the computed BIS variations are too small to be detected by instruments with a resolution power lower than 70,000. The computed RV variations agree with the Donati et al. (2011) observations. But the simulated  photometric variability is even lower than in the simulation by Hu\'elamo et al. (2008) with less than 1\%. While the spot may have evolved between the different observations, we can rather conclude that a simple spot model is not sufficient to explain all variables, and a  combination of a dark spot and bright accretion probably explains the observables of this young star. We note that a hot accretion spot should be seen as a continuum locally in the stellar photosphere and may also induce RV drift. However, because hot spots are usually smaller than cold ones, we expect their impact to be significantly weaker.

\begin{table}
  \centering 
  \caption{Input parameters for the simulation of TWHya }
  \label{donati}
\begin{tabular}{lll}
\hline
\hline
Parameter   &  Value        &       Ref.   \\
\hline
\hline
  grid      & 300      &     \\
   nrho     &  20     &          \\
\hline

limb   &   0.6       &      \\
radius [R$_{\odot}$] &  1.1           &  Donati et al. 2011  \\
Prot [day]   &  3.56 &  Setiawan et al. 2008\\
i [$^{\circ}$]       &  15   &    Donati et al. 2011  \\
gamma  [kms$^{-1}$] & 0. & \\
psi & 0. & \\
\hline

lat [$^{\circ}$]  & 80 &  \\
bright & 0. & fixed \\
size  [R$_{\star}$] & 0.14 &  \\
\hline

sigma0 [kms$^{-1}$] & 4.6 & ESPaDOnS resolution \\
window [kms$^{-1}$] & 20 & \\
step [kms$^{-1}$] & 0.1 & \\
\hline            
\hline
\end{tabular}
\end{table}
	
	\subsubsection*{HD189733}	
	
	HD189733 is an active star that hosts a transiting hot-Jupiter. Boisse et al. (2009) reported $\sim$2-months observations performed with the SOPHIE spectrograph simultaneously to photometric MOST data to monitor the stellar activity. We used their orbital solution to remove the effect of the hot Jupiter in the RV data and modeled the residuals. The simulation was performed with one dark spot and the input parameters reported in Table~\ref{boisse}. We chose a linear limb-darkening value of 0.8 according to Claret (2004) for [M/H]=0, T$_{\rm eff}$=5000K and $\log$g=4.5. In Fig.7 of Boisse et al. (2009), we observed a plateau in the variations of the parameters as a function of the rotational phase. This plateau is expected when the main spot is out of sight. We obtain a semi-amplitude close to the observed variations with a spot at 70$^{\circ}$ latitude and size of about 2\% of the stellar disk, $\Delta$RV/2=14ms$^{-1}$,  $\Delta$BIS/2=6ms$^{-1}$  and $\Delta$Flux of 1\%. The derived size of the spot agrees with the value estimated by HST observations of the star by Pont et al. (2007), thanks to occultations of spots by the transiting planet. Even if we can expect several spots in the photosphere of this star (e.g. Pont et al. 2007; Lanza et al. 2011), we observed that a one-spot model can well explain the observed RV, BIS, and flux variations (Fig.~\ref{FigHD189733}).

\begin{table}
  \centering 
  \caption{Input parameters for the simulation of HD189733 }
  \label{boisse}
\begin{tabular}{lll}
\hline
\hline
Parameter   &  Value        &       Ref.   \\
\hline
\hline
  grid      & 300      &     \\
   nrho     &  20     &          \\
\hline

limb   &   0.8       &   K2V star   \\
radius [R$_{\odot}$] &  0.766     &  Triaud et al. 2009  \\
Prot [day]   &  11.95 & Henry \& Winn 2008 \\
i [$^{\circ}$]       &  85.7   & Pont et al. 2007    \\
gamma  [kms$^{-1}$] & 0. & \\
psi & 0. & \\
\hline

lat [$^{\circ}$]  & 70 &  \\
bright & 0. & fixed \\
size  [R$_{\star}$] & 0.14 &  \\
\hline

sigma0 [kms$^{-1}$] &2.68 & SOPHIE resolution \\
window [kms$^{-1}$] & 20 & \\
step [kms$^{-1}$] & 0.1 & \\
\hline            
\hline
\end{tabular}
\end{table}
	
   \begin{figure}
   \centering
   \includegraphics[width=8cm]{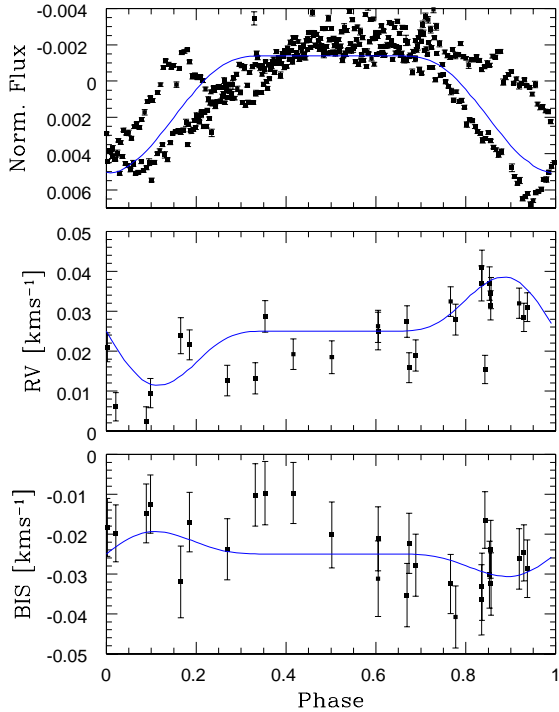}
      \caption{In blue lines, SOAP simulation of a star with the same characteristics as HD189733 observed with SOPHIE with a spot of 2\% of the stellar surface to reproduce the RV, BIS, and flux variations observed by Boisse et al. (2009) (black data points). 
             }
         \label{FigHD189733}
   \end{figure}

\section{Perspectives}

At this stage, and as presented in this paper, SOAP is at a quite basic model level. Because it is fast and clearly coded, improvements and developments can be easily implemented. Dumusque et al. (2011) already used this tool to simulate the evolution of the number of spots along the magnetic cycle of the Sun, and inferred the corresponding RV amplitude. These authors deduced a method to average the effect of the stellar activity to search for low-mass planets.

Additional improvements will be made on the tool.  SOAP simulates the photometric effect due to inhomogeneities on the stellar surface. However, it does not consider the effect that they also block the convection pattern. Moreover, spots and plages are not only areas darker or brighter, but have a different temperature than the stellar photosphere. They then present different spectra than the stellar photosphere, which could be seen as different CCF parameters. This wavelength dependency must be taken into account because, if most of the accurate spectrographs are now in the optical, soon near-infrared instruments will be available, a wavelength domain in which the amplitude of RV variation due to activity is supposed to be smaller (e.g. Hu\'elamo et al. 2008; Prato et al. 2008; Figueira et al. 2010; Crockett et al. 2011; Mahmud et al. 2011).
 
Stellar magnetic fields might be monitored by polarimetry of the stellar lines induced by Zeeman effect (Donati \& Landstreet, 2009) with instruments such as ESPADONS@CFHT, Narval@TBL, HARPSPol@ESO, or soon SPIRou@CFHT also dedicated to the search of planets around low-mass stars. A first attempt to look for relations between RV variations and polarized light is promising (Delfosse et al., in prep.). 
 SOAP could then be developed to simulate the polarimetric emission of the active star.

When the planets are transiting in front of their parent stars, their radii are measured. And when they are also characterized in RV, which yields  the mass, it leads to a measure of the mean density, which is the first step in the internal characterization of exoplanets. But the stellar inhomogeneities may have impacts on the radius determination derived from transit detections. One of the impacts is that it hampers a correct determination of the radius. 
That will be crucial for determining atmospheric components that are derived from the radius variations as a function of wavelength (e.g. Czesla et al. 2009).\\

Dedicated observations of active stars are therefore needed, if possible simultaneously, in RV, photometry, and polarimetry to characterize the different level of stellar variability, to understand the correlation between the different proxies, and to constrain the simulations. The release of thousands of light curves from the space missions CoRoT and Kepler dedicated to asteroseismology and the search for planets via transit will help us to understand statistically the photometric variability of these stars.

This tool is useful for computing and testing the effect of stellar activity on RV and photometry. It will help to understand the challenges related to the knowledge of stellar activity for the next decade: detect rocky planets in the habitable zone of their stars (from G to M dwarfs), understand the activity in the low-mass end of M dwarf (on which future projects such as SPIRou or CARMENES will focus), limitation to the sum of different transit observations to characterize the atmospheric components (from the ground or with Spitzer and JWST), and search for planets around young stars. These can be simulated with SOAP to search for indices and corrections of the effect of activity.

\begin{acknowledgements}
We are grateful to the referee for constructive comments. 
IB and NCS acknowledge the support by the European Research Council/European Community under the FP7 through a Starting Grant, as well from Funda\c{c}\~ao para a Ci\^encia e a Tecnologia (FCT), Portugal, through a Ci\^encia\,2007 contract funded by FCT/MCTES (Portugal) and POPH/FSE (EC), and in the form of grants reference PTDC/CTE-AST/098528/2008, PTDC/CTE-AST/098604/2008, and SFRH/BPD/81084/2011. 
\end{acknowledgements}


\begin{thebibliography}{}

\bibitem[aigrain]{aigrain}
Aigrain, S., Pont, F., and Zucker, S. 2012, \mnras, 419, 3147A

\bibitem[1996]{baranne}
Baranne, A., Queloz, D., Mayor, M., et al. 1996, \aaps, 119, 373

\bibitem[2009]{boisse09} 
Boisse, I., Moutou, C., Vidal-Madjar, A., et al. 2009, \aap, 495, 959

\bibitem[2010]{boisse10}
Boisse, I., Eggenberger, A., Santos, N.C. et al. 2010, \aap, 523, 88B

\bibitem[2011]{boisse11}
Boisse, I., Bouchy, F., H\'ebrard, G. et al. 2011, \aap, 528, A4

\bibitem[2007]{bonfils}
Bonfils, X., Mayor, M., Delfosse, X. et al. 2007, \aap, 474, 293

\bibitem[2004]{claret}
Claret, A. 2004, \aap, 428, 1001

\bibitem[2011]{crockett}
Crockett, C.J., Mahmud, N.I., Prato, L. et al. 2011, \apj, 735, 78

\bibitem[czesla]{czesla}
Czesla, S., Huber, K.F., Wolter, U., Schr\"oter, S. \& Schmitt, J.H.M.M. 2009, \aap, 505, 1277

\bibitem[desort]{desort}
Desort, M., Lagrange, A.-M., Galland, F., Udry, S. and Mayor, M. 2007, \aap, 473, 983

\bibitem[donati]{donati}
Donati, J.-F. \& Landstreet, J.D. 2009, ARA\&A, 47, 333D

\bibitem[donati11]{donati11}
Donati, J.-F., Gregory, S.G., Alencar, S.H.P. et al. 2011, \mnras, 417, 472D

\bibitem[dumusqeu]{dumuue}
Dumusque, X., Santos, N.C., Udry, S., Lovis, C., Bonfils, X. 2011, \aap, 527, 82D

\bibitem[2010]{figueira}
Figueira, P., Marmier, M., Bonfils, X. et al. 2010, \aap, 511, 55

\bibitem[hatzes]{hatzes}
Hatzes, A.P. 1999, ASPC, 185, 259

\bibitem[hatzes2]{hatzes2}
Hatzes, A.P. 2002, AN, 323, 392

\bibitem[henry]{henry}
Henry, G.W. \& Winn, J.N. 2008, AJ, 135, 68 

\bibitem[2008]{huelamo}
Hu\'elamo, N., Figueira, P., Bonfils, X. et al. 2008, \aap, 489, L9

\bibitem[huerta]{huerta}
Huerta, M., Johns-Krull, C.M., Prato, L., Hartigan, P. \& Jaffe, D.T. 2008, 678, 472

\bibitem[lanza]{lanza}
Lanza, A.F., Bonomo, A.S., Moutou C. et al. 2010, \aap, 520, A53

\bibitem[lanza2]{lanza2}
Lanza, A.F., Boisse, I., Bouchy, F., Bonomo, A.S. and Moutou, C. 2011, \aap, 533, 44

\bibitem[2011]{lovis}
Lovis, C., Dumusque, X., Santos, N.C. et al., arXiv:1107.5325

\bibitem[2011]{Mahmud}
Mahmud N.I., Crockett, C.J., Johns-Krull, C.M. et al. 2011, \apj, 736, 123

\bibitem[mayor]{mayor}
Mayor, M. \& Queloz, D. 1995, \nat, 378, 355

\bibitem[meunier]{meunier}
Meunier, N., Desort, M., Lagrange, A.-M. 2010, \aap, 512, 39M      

\bibitem[2002]{pepe02} 
Pepe, F., Mayor, M., Galland, F., et al. 2002, \aap, 388, 632

\bibitem[2007]{pont}
Pont, F., Gilliland, R.L., Moutou, C. et al. 2007, \aap, 476, 1347

\bibitem[2008]{prato}
Prato, L., Huerta, M., Johns-Krull, C.M. et al. 2008, \apj, 687, L103

\bibitem[1998]{queloz98}
Queloz, D., Allain, S., Mermilliod, J.-C., Bouvier, J. and Mayor, M. 1998, \aap, 335, 183

\bibitem[2001]{queloz}
Queloz, D., Henry, G.W., Sivan, J.P. et al. 2001, \aap, 379, 279
     
\bibitem[2009]{queloz09}
Queloz, D., Bouchy, F., Moutou, C. et al. 2009, \aap, 506, 303     
     
\bibitem[saar]{saar} 
Saar, S.H. \& Donahue, R.A.  1997, \apj, 485, 3195    

\bibitem[2003]{santos03}
Santos, N.C., Udry, S., Mayor, M. et al. 2003, \aap, 406, 373

\bibitem[201b]{santos10}
Santos, N.C., Gomes da Silva, J., Lovis, C. and Melo, C. 2010, \aap, 511, 54S

\bibitem[setiawan]{setiawan}
Setiawan, J., Henning, T., Launhardt, R. et al. 2008, \nat, 451, 38

\bibitem[takeda]{takeda}
Takeda, G., Ford, E.B., Sills, A. et al. 2007, ApJS, 168, 297

\bibitem[triaud]{triaud}
Triaud, A.H.M.J., Queloz, D., Bouchy, F. et al. 2009, \aap, 506, 377

\end{thebibliography}
\end{document}